# NEED FOR CRITICAL CYBER DEFENCE, SECURITY STRATEGY AND PRIVACY POLICY IN BANGLADESH - HYPE OR REALITY?

AKM Bahalul Haque

Depart of Electrical and Computer Engineering
North South University, Dhaka, Bangladesh

## ABSTRACT

*Cyber security is one of the burning issues in modern world. Increased IT infrastructure has given rise to enormous chances of security breach. Bangladesh being a relatively new member of cyber security arena has its own demand and appeal. Digitalization is happening in Bangladesh for last few years at an appreciable rate. People are being connected to the worldwide web community with their smart devices. These devices have their own vulnerability issues as well as the data shared over the internet has a very good chances of getting breached. Common vulnerability issues like infecting the device with malware, Trojan, virus are on the rise. Moreover, a lack of proper cyber security policy and strategy might make the existing situation at the vulnerable edge of tipping point. Hence the upcoming new infrastructures will be at a greater risk if the issues are not dealt with at an early age. In this paper common vulnerability issues including their recent attacks on cyber space of Bangladesh, cyber security strategy and need for data privacy policy is discussed and analysed briefly.*

## 1. INTRODUCTION

Bangladesh has one of the fastest growing internet users in the world. The number of internet users is increasing day by day very rapidly. In Bangladesh, a lot of people uses a smartphone. Mobile operators have taken this opportunity and used it as their touchstone. The operators worked on making the internet available to all the users at an affordable condition. Hence the flourish of mobile internet has given rise to the number of internet users. The number of Internet Service provider is also on the rise. They provide the users with high-speed internet all over the country. Almost all the universities are equipped with Wireless internet throughout the campus. Even there are a few public Wi-Fi hotspots.

All the factors have facilitated towards the number of increased internet users in our country. A study shows that till December 2018 almost 91 million people of Bangladesh are internet subscribers. Among them, 86 million are mobile internet users [1]. So, it is very much imaginable, mobile internet has created a boom in online service usage. According to Wikipedia, there are almost 90 million smartphone users in Bangladesh. This smartphone consumes a huge amount of data and people are constantly using a lot of services like photo sharing, social networking, file sharing etc. Mobile banking has also become very common. Banks are also providing their services over the internet to their customers through mobile applications. People are using ride-sharing application, online shopping etc and all other types of activities.

The usage of online activities is producing events and logs. These events are producing data. The online activity is making user's identity and signature open to the internet. The task that we are doing, the file that we are sharing as long as it is over the internet they can be breached. If there is data breach on a massive scale due to the lack of cyber security infrastructure that would bring in some very catastrophic consequences.

In this study, the present condition of Bangladesh with respect to cyber infrastructure, internet users, data breach, cyber security strategy and possible solutions will be discussed.

## 2. CYBER THREAT (COMMON VULNERABILITIES) ANALYSIS IN BANGLADESH

As the number of internet user increases, the threats towards the cyberspace increases with it. According to the security report of Kaspersky in 2015, it is seen that Bangladesh places second among all the other countries in the field of malicious infection. The number of unique users who are prone to cyber-attack is 69.55%. For the most part, these attacks are spam attacks. Almost 80% of them are spam attacks says Trend Micro Global Spam map. Moreover, a recent study by Bangladesh Computer Council showed that a huge number of unique IP addresses which were infected belongs to the popular mobile network operators. The number of IP address is approximately 34552. [2]

Figure 1 depicts a comparative scenario of CCM in Bangladesh and the rest of the world during the year 2015. CCM elaborates as Computers Cleaned per Mile. This is an infection metrics. It defines how many computers are cleaned per 1000 unique computers through MSRT tool. MSRT which refers to Malicious software removal tool is a free tool provided by Microsoft which removes potentially malicious software from the computer. It is seen that the encounter rate is significantly larger than that of the whole world.

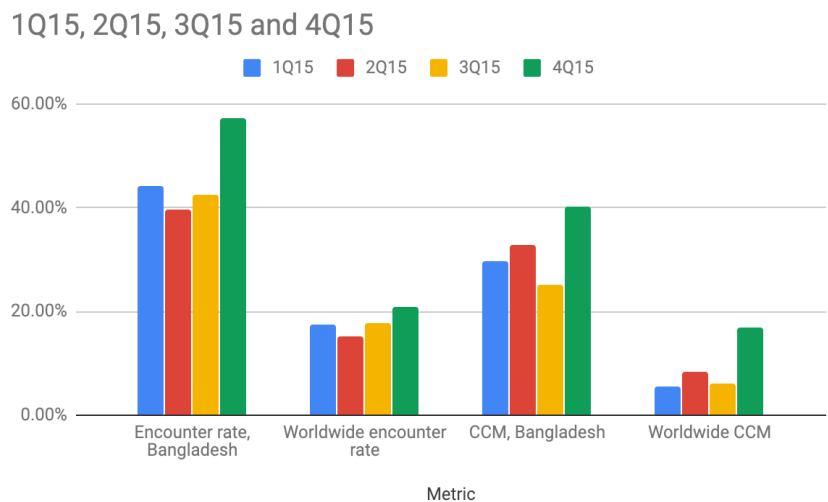

Figure 1: CCM Encounter Comparison between Bangladesh and the rest of the World

In the following figure (Figure 2), it is seen that most common malware in Bangladesh during the year 2015 was Virus, Trojan, and worm. All of them increased from 3rd to 4th quarter of the same year. The percentage of worm infected computers increased from 21.1 to 29.6 percent, the percentage increased from 19.4 to 22.9 in case of trojan and finally the percentage of virus-infected computers increased from 6.3 to 9 percentage during 3rd and 4th quarter respectively.

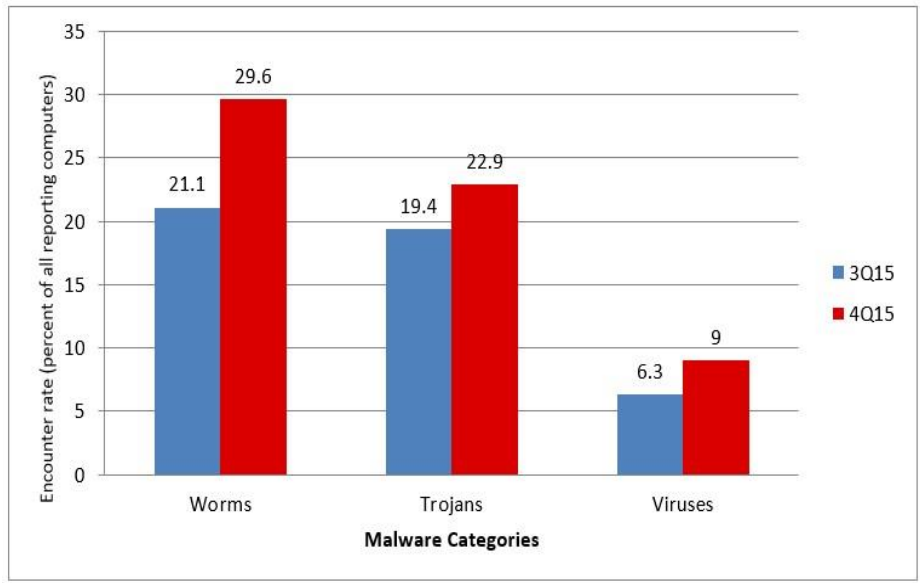

Figure 2: Malware Categories and infected computers

Figure 3 depicts another graph showing the malware families encountered during 4Q15 . In this diagram we can see that Win32/Ippedo and Win32/Gamarue were most common in Worm and infected a large number of computers. In case of Virus it was Win32/Ramnit and in Trojan category it was Win32/Skeeyah.

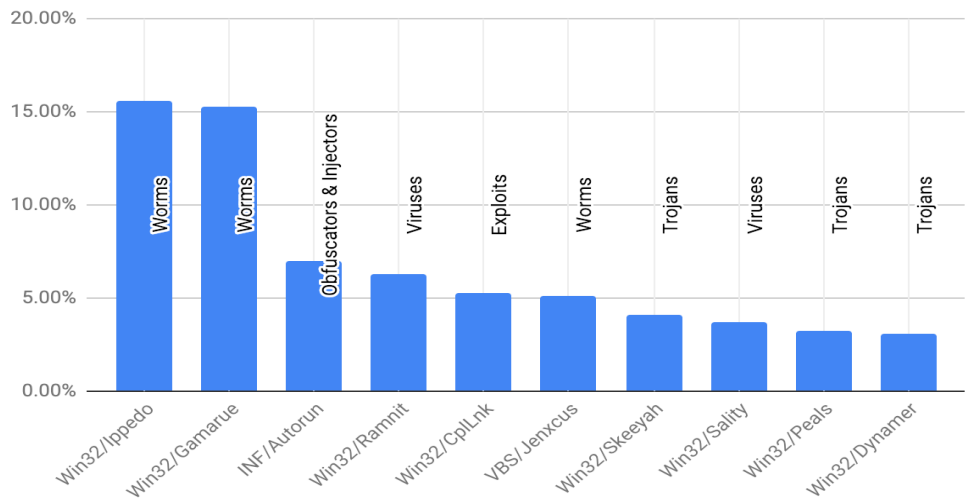

Figure 3: Most common Malware categories detected in 4Q15

While considering the category of potentially unwanted software, browser modifiers, and software bundlers were most significant and alarming. In the following table (Table 1) the percentage of computers infected by browser modifiers and software bundlers are shown during 4Q15.

Table 1: Statistics of Unwanted Software Threats in 4Q15

| Family | Most Significant Category | % of reporting computers |
|--------|---------------------------|--------------------------|
|        |                           |                          |

| Win32/ Diplugem | Browser Modifiers | 6.00% |
|---|---|---|
| Win32/ SupTab | Browser Modifiers | 6.00% |
| Win32/Dowadmin | Software Bundlers | 5.70% |
| Win32/OutBrowse | Software Bundlers | 3.50% |
| Win32/Mizenota | Software Bundlers | 3.40% |

## 3. CYBER THREATS IN THE FINANCIAL SECTOR AND MILITARY DOMAIN

There has been a conflicting situation a rising in the military domain due to the advancement of internet and computer technology. Due to this rapid advancement of cyber technologies, large organizations and corporations have turned into a potential target for cyber-attacks, losing billions of important piece of information and confidential data raising doubts about the nation's stability. [3]

Cyber-attacks have been a recurring occurrence making the news headline in the recent time. These attacks take place when an adversary intends to intervene an electronic infrastructure with the financial motive of gaining benefit from them and thus attempts to read, alter, delete or even add information from an electronic infrastructure. There has been an indication of many nations supplying cyber facilities to military entities negotiating over specific benefits in exchange that is to overtake particular regions. The widespread Internet access has facilitated the terrorist organizations to communicate over an ample number of platforms.

Thus it has become a vital issue of concern for the government to ensure and implement proper security strategies throughout the nation to protect the nation from all kinds of approaching cyber-attacks and take measures accordingly by stabilizing the nation's security and military tactics.

Bangladesh has encountered quite a few security challenges in the recent time, and the trend is continuing due to the escalation of cyber technologies. There have been multiple incidents of state-sponsored cyber-attacks in the Bangladesh National Defense College and Bangladesh Bank which has stirred and questioned the credibility and reliability of the military organization of the nation creating a big hollow in the country's safety. State-affiliated adversaries often involve many targeted objectives taking an interest either in the political or military system of the country. Due to cyber theft, Bangladesh Bank had to face a loss of almost millions of dollars and was left susceptible to further attacks by the same nation. The network breaches relating to criminal and terrorist organizations have adopted the means best suited to their intention. They have got in hand many cheap and accessible software as an effective means to take down the security protocols and hamper the credibility of our military administrations.

The UN charter, however, stated to the incident of Burmese hackers taking over the NDC website that Bangladesh should have counter attacked as self-defense towards this digital armed attack but instead they have shown constant self-restraint to such incitement and gained some noteworthy praise from the international community. This, however, also demonstrates to some extent the ignorance and incapability of Bangladesh militaries.

Cyberspace goes far beyond geographical locations, and potential adversaries are widely spread for cyber-attacks starting from a mere hacker to a well-organized state-sponsored cyber-attacks. State-affiliated attacks often have the objectives of gaining Intel about political, commercial or military organizations of the targeted nations. To identify an intruder in the cyberspace, it may require exhaustive forensic research.

Bangladesh being a developing country, is yet to go a long way to understand the intensity of a country's cybersecurity system and the lack of awareness can have detrimental impacts on the national security system of Bangladesh. The incident of Holy Artisan Bakery in 2016 is a nightmare that still terrorizes the whole nation. That incident can be an apparent demonstration of the fact that the Bangladesh military and security intelligence was oblivious of the communication terms of the attackers, even though on the internet there were deliberate entrails of their activities through fund transfer from Canada and UK. Unfortunately, this shows how feeble our security intelligence teams were regarding identifying the attackers and preventing a terrorist attack taking place beforehand.

This indeed is a high time for Bangladesh armed forces to take strict initiatives of strengthening and ensuring the proper placement of national security systems. As can be seen, cyber-attacks have risen to a level surpassing any physical attacks, we should take in consideration of any and every factor necessary for the establishment of the national security defense system that could be improved to fill up the void in cyberspace. Being able to detect potential threats and cyber vulnerabilities beforehand and ensuring the successful execution of military mission and proper establishment of expeditionary force should be the primary concern for Bangladesh. This can be accomplished through the approach of appropriate strategic planning and abiding by specific policies relating to cybersecurity.

Bangladesh military holding to its national objective of the defense of country is dependent on countries like China for equipment support such as computer and network related components is mostly imported from China which automatically pose a threat to the nation since the manufacturer country is well aware of their components and thus have an upper hand on the military defense system of our country.

Thus to safeguard our nation's security, we should be able to acknowledge the rapid increase of cyber threats and come up with possible approaches to retaliate as it is said prevention is always better than cure. Our knowledge is the dominant key that must be utilized to its level best for the better perception of our cyber-security. A useful framework for cyberspace would include cyber defense, cyber intelligence, and cyber counter-offensive. This leads to the conclusion that Bangladesh should develop specific strategies and concepts for the recruitment process and the means of spreading awareness through training and education should be tactical to not leave any vulnerable strings behind. Bangladesh has made allies with the militarily developed countries like Australia and has signed an MOU on December 24, 2008, to share Intel and intelligence on counter-terrorism and later on 2015 and MOU on police to police cooperation.

# 4. SURVEY RESULT ANALYSIS OF CYBER DEFENCE STRATEGY IN BANGLADESH

A survey was prepared among the internet users of Bangladesh. The respondents were people from various kinds of profession. They were the day to day internet users. They were asked a few questions regarding the cyber defense situation, cyber defense strategy of Bangladesh. The result of the study is given below –

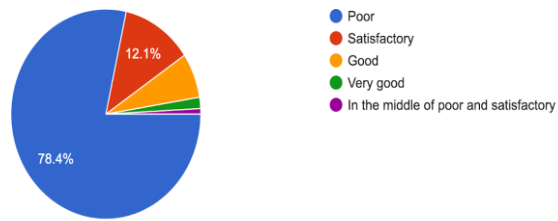

Figure 4: Survey of Cyber Defense situation of Bangladesh

Common people using various internet services in their day to day life think that the present cyber defense situation is not up to the mark. 78.4 % of them opines that the condition is poor. Only less than 2% people think the condition is very good. It might be due to the recent event of Bangladesh Bank or the attack on NDC. Nowadays social media has significance in our life and people's opinion and awareness vary a lot since the social networking sites have flourished and internet has become available. These sites spread news faster than any other news media around us and let people think and analyse the situation. Common people are becoming more aware about themselves and the media they are using.

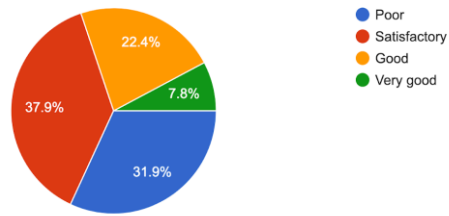

Figure 5 : Survey of People's' Awareness level of Cyber Security

Opinion about the awareness level of the internet users have varied. It is seen in Figure 5, most of the respondents (37.9%) think that their awareness level is satisfactory and almost 31.9% think that they have poor awareness level about cyber security. 22.4% and 7.8% of the respondents have opined that their awareness level is "Good" and "Very Good" respectively. In case of this people's understanding, it is assumed that the sense of " thinking we know everything " might have affected the study a bit , but still a very good number of respondents do not have enough knowledge about the cyber security which is a very sensitive issue as Bangladesh is entering an era of digital age.

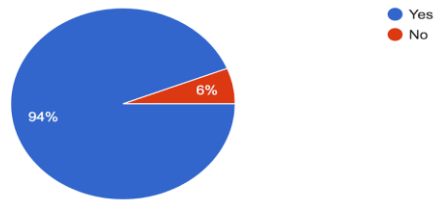

Figure 6: Survey on Recommendation about making common people aware about cyber-attacks.

Figure 6 depicts users' opinion about making the common mass aware about cyber security. As we know that, awareness rings change to the society, community and to the country. In light of this, awareness about security infrastructure and about the usage of internet can go a long way to protect users while surfing the internet for day to day purpose.

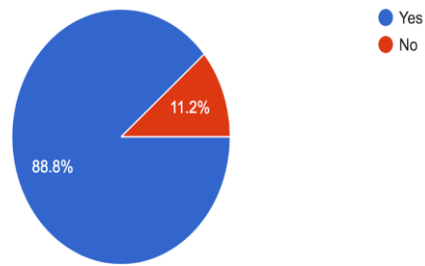

Figure 7 : Survey Result of the Demand of Cyber Defense Strategy in Bangladesh

To protect the cyberspace of any country it needs to have proper policy management and implementation. The online activities make people exposed to the outside world. While using internet the infrastructure on which the cyberspace is built on should be well shielded for the sake of protecting privacy. Moreover, if the infrastructure is weak, the national security is also at stake. The survey shows (Figure 7) that, 88.8% of the total respondents think that, Bangladesh should have a strong Cyber Defense Strategy immediately.
.

## 5. RECOMMENDATIONS FOR IMPROVING CYBER DEFENSE OF BANGLADESH

The way the adversaries are taking place it shows us how sophistically equipped the cyber attackers are to drastically disrupt the cyber world if we are to perceive the broader picture, we get the idea of the cyber attackers getting powerful every day with very little successful counterattacks. This indeed is creating havoc throughout the nation.

So the defense authorities of a nation must take the appropriate measures to combat all kinds of cyber-attacks adopting proper strategies, plans and policies to deter cyber-attacks at every level starting from the Strategic and Operational level. Establish cyber infrastructure for the militaries to oversee and manage proper security protocols. Acquiring sufficient knowledge and sophisticated equipment to increase the capabilities of cyber security intelligence and security defense organizations. Execution of critical missions should be facilitated by proper terms and conditions with several other agencies.

A blockchain is a distributed peer to peer networks transaction which heavily relies on the transparency of communication to safeguard data.[4] A distributed network makes it much more difficult for cyber-attacks to take place since the networks is publicly visible and is witnessed by everyone. Thus making a network almost impossible to hack by cyber attackers.

In cryptography and security, blockchain has quite made its foundation through decades of research [5]. The concept is being used for the enhancement of cyber security against cyber-attacks. To sustain against the streams of cyber threats and attacks, block chain could be an effective solution since it is a decentralized structure there is no single point of attack thus it would never allow excessive requests to pass through making it impossible for the infrastructure to fail. Blockchain has the infrastructure of a bitcoin crypto currency. Blockchain being a centralized structure can prevent data manipulation and fraud since it is immutable in nature.[6] Decentralization and immutability are the most important characteristics of Bitcoin which makes the network secure from cyber-attacks [7].

## 6. NATIONAL CYBER SECURITY STRATEGY OF BANGLADESH

According to the gadget published from Information Technology Division, Ministry of Posts Telecommunications and Information Technology, Bangladesh in March 2014, The National Cyber Security Strategy of Bangladesh has been aligned with the ITU (International Telecommunication Union) [8]. This strategy is a long term process. It aims to address and remove the threats risks and mitigation strategies in legal, technical and organizational level. Bangladesh has entered the digital era now. The government has been trying its best to digitize the country in every sector. Now, there is a central database of all the citizens of our country including the fingerprint and retinal scan. Mobile operators are collecting user data including the NID (National Identification) and fingerprint. Banks are also given access to the APIs for data processing. All the factors have given rise to the fact that protecting the cyberspace of Bangladesh has been more important than ever now. According to the aforementioned gadget, apart from breaching confidentiality, integrity and authenticity of regular communication the cyber threats towards the cyber defense of our country are the usage of political and industrial espionage to obtain illicit information, insecurity towards the government due to borderless anonymous communication, manipulation of smart grid technologies, disclosing information which is of importance to the National Security etc.

Priorities of National Cyber Security Strategy should have three national priorities:

• Legal Measures;

• Technical and Procedural Measures; and

• Organizational Structures.

### 6.1. Priority One - Legal Measures

Legal measures have two actions, respectively Cybercrime Legislation and Government Legal Authority. A legislative committee has to be formed regarding the cyber security strategy. The cybercrime law has to be set up in line with the ITU Policy. The law will be overseen and analyzed by the Ministry and the legislative committee. Anyone with other groups and organizations from private sectors can take part in this process so that the law can be better equipped with both the public and private sectors. Cyber security being a global concern, calls for the need of unified correlation between the national and regional cybercrime legislation in order to scrutinize the possible cybercrimes. The government will have the legal authority to ensure the security of the cyberspace for mass people. To protect the cyberspace and to protect the cyber infrastructure of the country the government can make cyber security division, cyber security response team and supervise these teams. Moreover, they should also promote awareness and training for people. The government will solicit plans, negotiate with other parties to solve and plan long term security solutions to alleviate the cyber infrastructure of our country.

Technical and Organizational Measures includes structuring of cyber security framework at national and regional level. The cyber security framework of our country should define strong and efficient policies at the organizational level to ensure the goals of the strategy are met correctly. The goals of the strategy include assurance and security of information, proper risk management, managing organizational assets and ensuring both environmental and physical safety.

## 6.2. Priority Two - Technical and Procedural Measures

Bangladesh National Cyber security framework focus on the compulsory security standards that all the stakeholders of an organization or corporation should maintain in order to have compliance with the minimum level of law enforcement required for the state. The goals of the strategy include assurance and security of information, proper risk management, scrutinizing employees while recruiting, managing organizational assets and ensuring both environmental and physical safety.
Protection of critical information infrastructure is of utmost importance to the government which can be done through spreading awareness and adopting appropriate security measures through several amendments in the official service contract beforehand to uphold a secured cyberspace. The government should further classify sensitive data and ensure proper cybersecurity risk management schemes in the cyber systems. The capacity, capability, and jurisdiction of the law enforcement agencies have to be increased to better protect the cyberspace of our country.. Increasing capability and capacity can be achieved through proper research and development activities. Universities and other public and private research organizations can play a vital role in participating in the research and development activities.

Weak infrastructure is what creates the basis for the strong awareness of cyber security. Thus identify vulnerabilities and threats in the system is very important to preserve the integrity of confidential information and avoid the exploitation of sensitive data from cyber-attacks. The necessary actions that must be taken for reducing this threats include a means of system assessment to comprehend the potential risks of data breaching and the corresponding consequences that follows. The sensitive infrastructure must be protected through schemes such that to get hold to the system a penetration test must be obligatory.

## 6.3. Priority Three - Organizational Structures

Organizational Structures include the cyber security role of government, National Cyber Security Council, National Incident Management Capacity, Public-Private Partnerships, Cyber security Skills and Training and National Culture of Cyber security.

In order to combat cyber-attacks, contribution and cooperation from every citizen is important. Government plays the core responsibility of security for the betterment of national citizens. Enhancement of cyber security schemes requires the government to reinforce its own national infrastructure so that it can help the public with the required services. A National Cyber Security coordinator is often addressed who has direct contact with the Head of Government and thus have access to all the necessary resources to carry out major inter-government activities to safeguard the national security information infrastructure at the strategic level. National Security Council is the central point for both the public and private sector coordination and interaction on cyber threats. The roles carried out by the council includes a thorough plan to protect the critical infrastructure and provide ample guidance to the organizations holding critical infrastructures such as banking, telecommunication etc. At times of severe national cyber-attacks on security infrastructures, the council's capacity to response is also measured. Consequently the council further works hand in hand with the government and intelligence agencies to detect security vulnerabilities and the possible counterattacks. For further assistance they engage in several research and development work for the escalation of cyber security technologies. The National Security Council have also aligned in collaboration with the International Multilateral Partnership against Cyber Threats (ITMPAC).

In order to respond immediately at critical cyber-attack incidents, government should have a continuous evaluation of the cyber security events along with its strategic solutions which can be effectively achieved through association with the private-public organizations both nationally and globally.

Private sectors owns the critical infrastructures and have better expertise and knowledge on this aspect of cyber security development and maintenance. Thus, government must comply with the private sectors to alleviate the national security issues by the procurement of adept professionals. This enables the government a strategic gain of international cooperation in the necessary areas.

Along with spreading awareness, a crucial initiative is to properly train and educate the cyber security professionals at several different levels. It will make them capable of identifying vulnerabilities at the critical infrastructures and take appropriate measures to prevent cyber-attacks. Thus proper collaboration with global partners through research and training should be maintained to better protect the cyberspace.

Since the world is now ruled by technologies, there's always multiple alternatives to a potential cyber attacker. A lack of consciousness from either the administrators or the end users can imply serious impacts on the overall security infrastructure. So it is important for the government and the stakeholders to be vigilant and responsive at all time and take incentives through education, research, development and training.

## 7. NEED FOR DATA PROTECTION AND PRIVACY POLICY IN BANGLADESH

Privacy and data protection is one of the most important issues nowadays. Due to the advancement of modern technologies and internet, the presence of users over the internet has increased more than ever. Increased number of internet users share huge amount of data during day to day life while surfing the internet. Within this huge data there are personal data of the user also. Personal data refers to any identifiable that can identify a person physically or over the internet. It might include location history, interest hobbies, telephone number, bank information, email information etc. [9]. Then there comes the sensitive personal data which needs to be processed with special care and extra security. According to GDPR, sensitive personal data is any data that reveals information regarding anyone's

- racial or ethnic origin,
- political opinions,
- religious or philosophical beliefs,
- trade union membership,
- genetic data, biometric data,
- data concerning health
- Data concerning a natural person's sex life or sexual orientation.

According to data protection Act (DPA, 1998) sensitive personal data consists of

- the racial or ethnic origin of the data subject,
- political opinions,

- religious beliefs or other beliefs of a similar nature,
- whether the data subject is a member of a trade union,
- physical or mental health or condition,
- about sexual life,
- the commission or alleged commission by the data subject of any offence,
- Any proceedings for any offence committed or alleged to have been committed by the data subject, the disposal of such proceedings or the sentence of any court in such proceedings [10].

Data privacy and protection is like a new member in the club. People still are not aware of their data being taken away or data theft. There is no explicit data privacy and protection law of Bangladesh. In the constitution of Bangladesh there is a mention about the citizens' right to their own data though there is no explicit mention of "data privacy or data protection". In Part 3, Under the Fundamental Rights Section, Article 43, of the Constitution of the People's Republic of Bangladesh it written that "Every citizen shall have the right, subject to any reasonable restrictions imposed by law in the interests of the security of the State, public order, public morality or public health – (a) to be secured in his home against entry, search and seizure; and (b) to the privacy of his correspondence and other means of communication." In 2006, ICT Act and later recently Digital Security act 2018 was enacted in October, 2018 in the parliament. Digital Security Act, 2018 is more acute towards the criminals according to the several articles in the act. This act only deals with the crime happening with digital documents and data. It does not comply with any kind of activities done in case of any files and or records in paper format. The punishment is also severe for the criminals [11]. According to Official Secret Act 1923, any person who assists in cybercrimes without getting physically involved is equally hold guilty and will be punished accordingly. Hacking or intruding in the digital infrastructure and alteration and exploitation of information is considered to be a serious offence [12].

These are the only acts which are implemented in Bangladesh. These acts are for punishing the criminals who commit cybercrimes in the eyes of the law enforcement agencies on the basis of this law. The important issues, that has been taken into account is the Data Protection and Privacy Policy for our country. European Union has already implemented the General Data Protection Regulation (GDPR) in May, 2018 which is a historical achievement for the region. The citizens of that region are given control over their data, whether it is processed by home or abroad .As long as the citizen is from the European Union member countries, the data has to be collected, processed and analyzed according to the GDPR[13]. This outstanding privacy policy can be taken as an example and our citizens can be given a lot of control over their data and moreover, the government can utilize their data to better analyses the current situation of the country. There are quite a few organizations which have explicitly shown information online. They provide all kinds of necessary information like name, address, phone number, parents' names, educational institution etc. Here, in Figure 8, the information that has been shown is mine. I was involved with this specific organizations for almost 4 years, still the information can be retrieved from their website with a simple google search. Moreover, the information about other colleagues of mine can also be retrieved from this website very easily. These are one of the many examples of data leakage. This sensitive data can be used for identity theft as all the authentic information are shown publicly and anyone with a simple knowledge of internet browsing can have access to this information. Considering these acts, Bangladesh is in dire need of a strong data protection and privacy policy where everything regarding data collection, processing and storage will be well defined and the users should have better control over their data.

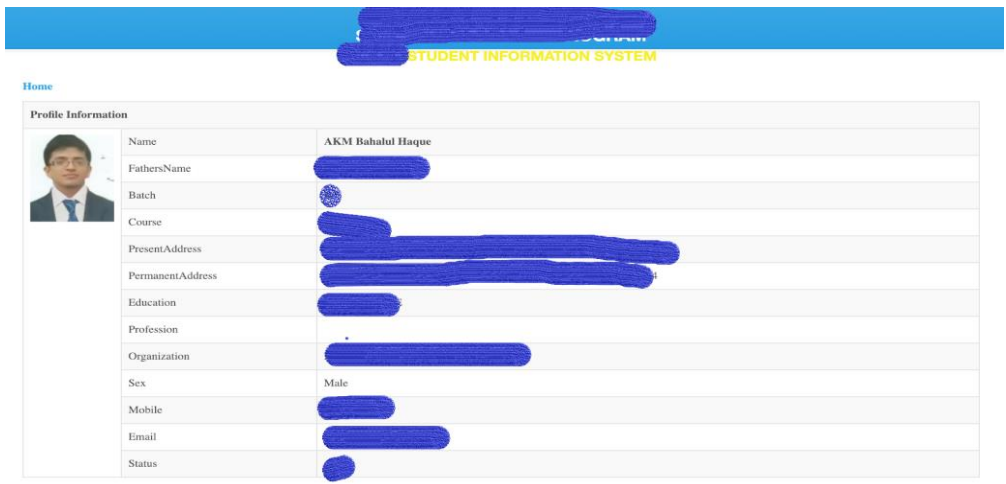

Figure 8 : Services revealing sensitive personal data

## 8. INITIATIVES OF BANGLADESH GOVERNMENT

Bangladesh is racing towards the infrastructural development. Moreover, due to the recent advancement of information and communication technology sector, the modern day web tracking and surveillance is also possible. There are a lot of tracking mechanisms [14]. Illegal web tracking can make the user vulnerable to a lot risks. Information can be easily get into the wrong hand. To prevent all these things government is trying to implement strong cyber security infrastructure and incident response team. Government has started initiative like Cyber Incident Response Team [14]. The task of this team is to

1. Protect the cyber security infrastructure of e-governance of Bangladesh

2. Respond to incidents as early as possible to improve the incident response management capability in case of any kind of security breach or event.

3. To promote the cyber security awareness and environment among the common people.

4. Provide security solutions to the national cyber infrastructures like data centers.

### 8.1. Improvised Initiatives

Government can consider establishing high tech specialized organization that is solely dedicated to cybercrime investigations. High end collaboration with international cyber defense agencies for maintaining an effective security framework. Systematic development of training and education for producing a skilled security workforce thus minimizing the gap between demand and supply of performing effective cyber security roles. Development of a scientific research center to support skilled people with necessary resources and also to gather new and innovative ideas for the effective implementation in the cyber domain. This would increase the participation of the people to fill up the gap of a lack of cyber and security knowledge. Every common citizen can have a clear vision of the fundamental cyber security knowledge and its importance to combat against potential cyber threats.

# 9. CONCLUSIONS

Cyber-space of Bangladesh has expanded over time with the advent of new technologies. People of all walks of life has their footsteps on the cyber arena which has significantly increased the importance of cyber security. The information we produce and share over the internet is vulnerable due to the lack of strong cyber defense infrastructure. Personal data is not protected and users do not have enough control over their own data. So, a strong cyber defense infrastructure and proper data protection regulation is a critical need for Bangladesh. Only then Bangladesh can take a step towards being a developed country and can be considered as a secured digital Bangladesh

## ACKNOWLEDGEMENTS

I would like to thank my friend and colleague Ms. Omanisha who has helped me throughout the journey of this research by motivating me.

**AUTHORS**

AKM Bahalul Haque is currently working a Lecturer of Department of ECE, North South University, Bashundhara, and Dhaka 1229. He has achieved is M.Sc. In Information Technology from Fachhochschule Kiel, Germany, 2018. He achieved his Bachelor of Science (Engineering) in Computer Science and telecommunication engineering in 2014. Has published two of his papers in International Journal. He specializes in Cyber Security, Cloud Computing, Data Privacy and protection. He has one-year experience as Security Engineer and one-year experience as Vulnerability Detection Consultant in Cyber Security Division.

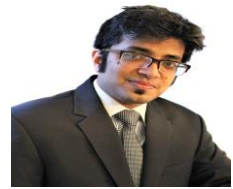